\documentclass{conm-p-l}

\input epsf

\copyrightinfo{2002}
  {American Mathematical Society}

\newtheorem{theorem}{Theorem}[section]
\newtheorem{lemma}[theorem]{Lemma}

\theoremstyle{definition}

\theoremstyle{remark}

\numberwithin{equation}{section}

\begin{document}

\title{The Large Radius Limit for Coherent States on Spheres}

\author{Brian C. Hall}
\address{Department of Mathematics, University of Notre Dame, Notre
Dame, IN 46556, USA}
%\curraddr{Department of Mathematics and Statistics,
%Case Western Reserve University, Cleveland, Ohio 43403}
\email{bhall@nd.edu}

\author{Jeffrey J. Mitchell}
\address{Department of Mathematics, Baylor University, Waco, 
TX 76798, USA}
\curraddr{Department of Mathematics, Louisiana State
University, Baton Rouge, LA 70803-4918, USA}
\email{Jeffrey\_Mitchell@baylor.edu}
\thanks{The second author was supported in part by NSF Grant DMS-9970882.}

%    General info
\date{April 18, 2002.}


%\keywords{Differential geometry, algebraic geometry}

%-----------------------------------------------------------------------

\begin{abstract}
This paper concerns the coherent states on spheres studied by the authors in
[J. Math. Phys. \textbf{43} (2002), 1211-1236]. We show that in the
odd-dimensional case the coherent states on the sphere approach the classical
Gaussian coherent states on Euclidean space as the radius of the sphere tends
to infinity.
\end{abstract}

\maketitle

\section{Introduction}

In our earlier work \cite{HM} we constructed coherent states and an associated
resolution of the identity for a quantum particle whose classical
configuration space is a $d$-dimensional sphere. Although the main result of
\cite{HM} is a special case of results of Stenzel \cite{St} (building on
results of Hall \cite{H1,H2}), we give a substantially different description
based on the ``complexifier'' approach of Thiemann \cite{T} and the ``polar
decomposition'' approach of Kowalski and Rembieli\'{n}ski \cite{KR1}. In
\cite{HM} we also give self-contained elementary proofs of all the main
results. See also \cite{H3} for a survey of related results.

We consider $S^{d},$ the sphere of radius $r$ in $\mathbb{R}^{d+1},$ viewed as
the configuration space for a classical particle. We consider also the
corresponding phase space, the cotangent bundle $T^{\ast}(S^{d}),$ which we
describe as
\[
T^{\ast}(S^{d})=\left\{  (\mathbf{x},\mathbf{p})|x^{2}=r^{2},\,\mathbf{x}%
\cdot\mathbf{p}=0\right\}  .
\]
Here $\mathbf{p}$ is the \textit{linear} momentum which must be tangent to the
sphere at $\mathbf{x},$ that is, orthogonal to $\mathbf{x}.$

We now briefly review the results of \cite{HM}. Following the complexifier
approach of Thiemann we first choose a constant $\omega$ with units of
frequency, whose significance will be discussed in Section 2. Then we consider
the ``complexifier'' function on $T^{\ast}(S^{d}),$ defined by
\[
\text{complexifier}=\frac{\text{kinetic energy}}{\omega}=\frac{p^{2}}%
{2m\omega}=\frac{j^{2}}{2m\omega r^{2}},
\]
where $j^{2}$ is the total angular momentum. Then we define complex-valued
functions $a_{1},\cdots,a_{d+1}$ on $T^{\ast}(S^{d})$ by the formula
\begin{align}
a_{k}  & =e^{i\left\{  \cdot,\text{complexifier}\right\}  }x_{k}\nonumber\\
& =\sum_{n=0}^{\infty}\left(  \frac{i}{2m\omega r^{2}}\right)  ^{n}\frac
{1}{n!}\underset{n}{\underbrace{\left\{  \cdots\left\{  \left\{  x_{k}%
,j^{2}\right\}  ,j^{2}\right\}  ,\cdots,j^{2}\right\}  }}%
\label{classical.complexifier}%
\end{align}
where $\left\{  \cdot,\cdot\right\}  $ is the Poisson bracket. A calculation
gives the explicit formula
\begin{equation}
\mathbf{a}\left(  \mathbf{x,p}\right)  =\cosh\left(  \frac{j}{m\omega r^{2}%
}\right)  \mathbf{x}+i\frac{r^{2}}{j}\sinh\left(  \frac{j}{m\omega r^{2}%
}\right)  \mathbf{p}\label{a.form}%
\end{equation}
The functions $a_{k}$ satisfy $\left\{  a_{k},a_{l}\right\}  =0$ and
$a_{1}^{2}+\cdots+a_{d+1}^{2}=r^{2}.$ The map $\left(  \mathbf{x,p}\right)
\rightarrow\mathbf{a}\left(  \mathbf{x,p}\right)  $ is a diffeomorphism of
$T^{\ast}(S^{d})$ with the \textit{complex sphere} $S_{\mathbb{C}}^{d},$
where
\[
S_{\mathbb{C}}^{d}=\left\{  \mathbf{a}\in\mathbb{C}^{d+1}|a_{1}^{2}%
+\cdots+a_{d+1}^{2}=r^{2}\right\}  .
\]

We now consider a quantum particle moving on the sphere. We take the quantum
Hilbert space to be the position Hilbert space $L^{2}(S^{d}).$ (See \cite{HM}
for a more abstract approach.) We consider the quantum complexifier given by
\[
\text{complexifier}=\frac{\text{kinetic energy}}{\omega}=\frac{J^{2}}{2m\omega
r^{2}},
\]
where $J^{2}$ is the total angular momentum operator given by
\[
J^{2}=-\hbar^{2}\sum_{k<l}\left(  x_{k}\,\frac{\partial}{\partial x_{l}}%
-x_{l}\frac{\partial}{\partial x_{k}}\right)  ^{2}.
\]
By analogy to (\ref{classical.complexifier}) (replacing the Poisson bracket
with the commutator divided by $i\hbar$) we define non-self-adjoint operators
$A_{k}$ by the formula
\begin{align}
A_{k}  & =e^{i[\cdot,\text{ complexifier}]/i\hbar}X_{k}\nonumber\\
& =\sum_{n=0}^{\infty}\left(  \frac{1}{2m\omega r^{2}\hbar}\right)  ^{n}%
\frac{1}{n!}\underset{n}{\underbrace{\left[  \cdots\left[  \left[  X_{k}%
,J^{2}\right]  ,J^{2}\right]  ,\cdots,J^{2}\right]  }}.\label{qa.intro}%
\end{align}

We call these operators the annihilation operators. There is also a polar
decomposition for the $A_{k}$'s (generalizing a formula of \cite{KR1} for the
$S^{2}$ case) and an explicit formula similar to (\ref{a.form}) but with
certain quantum corrections \cite[Eq. (38)]{HM}. We may also express $A_{k}$
as
\[
A_{k}=e^{-\tau\tilde{J}^{2}/2}X_{k}e^{\tau\tilde{J}^{2}/2}%
\]
where $\tilde{J}^{2}=J^{2}/\hbar^{2}$ is the dimensionless form of the angular
momentum operator and where $\tau$ is the dimensionless parameter given by

\[
\tau=\frac{\hbar}{m\omega r^{2}}.
\]
The $A_{k}$'s satisfy $\left[  A_{k},A_{l}\right]  =0$ and $A_{1}^{2}%
+\cdots+A_{d+1}^{2}=r^{2}I.$

We now define the coherent states to be the simultaneous eigenvectors of the
$A_{k}$'s. There is one coherent state for each point $\mathbf{a}$ in the
complex sphere $S_{\mathbb{C}}^{d},$ which means one for each point in the
classical phase space, since we identify $S_{\mathbb{C}}^{d}$ with $T^{\ast
}(S^{d})$ by means of (\ref{a.form}). The coherent states are given by the
heuristic expression
\[
\left|  \psi_{\mathbf{a}}\right\rangle =e^{-\tau\tilde{J}^{2}/2}\left|
\delta_{\mathbf{a}}\right\rangle ,\quad\mathbf{a}\in S_{\mathbb{C}}^{d}%
\]
where $\left|  \delta_{\mathbf{a}}\right\rangle $ is supposed to be a position
eigenvector satisfying $X_{k}\left|  \delta_{\mathbf{a}}\right\rangle
=a_{k}\left|  \delta_{\mathbf{a}}\right\rangle .$ For $\mathbf{a}$ in the real
sphere $\left|  \delta_{\mathbf{a}}\right\rangle $ is a generalized function
and $e^{-\tau\tilde{J}^{2}/2}\left|  \delta_{\mathbf{a}}\right\rangle $ is a
smooth function on $S^{d}.$ For $\mathbf{a}$ in the complex sphere $\left|
\psi_{\mathbf{a}}\right\rangle $ can be defined by analytic continuation with
respect to $\mathbf{a}.$ See \cite[Prop. 1]{HM}.

Explicitly we have
\[
\psi_{\mathbf{a}}\left(  \mathbf{x}\right)  =\rho_{\tau}^{d}(\mathbf{a}%
,\mathbf{x}),\quad\mathbf{x}\in S^{d},\,\mathbf{a}\in S_{\mathbb{C}}^{d}.
\]
where $\rho_{\tau}^{d}$ is the heat kernel on the $d$-sphere. Here $\rho
_{\tau}^{d}\left(  \mathbf{a,x}\right)  $ is initially defined for
$\mathbf{a}$ and $\mathbf{x}$ in $S^{d},$ but we can extend to $\mathbf{a}\in
S_{\mathbb{C}}^{d}$ by analytic continuation. For odd-dimensional spheres we
have the formulas
\begin{align}
\rho_{\tau}^{1}(\mathbf{a},\mathbf{x})  & =(2\pi\tau)^{-1/2}\sum_{n=-\infty
}^{\infty}e^{-(\theta-2\pi n)^{2}/2\tau}\nonumber\\
\rho_{\tau}^{3}(\mathbf{a},\mathbf{x})  & =(2\pi\tau)^{-3/2}e^{\tau/2}\frac
{1}{\sin\theta}\sum_{n=-\infty}^{\infty}(\theta-2\pi n)e^{-(\theta-2\pi
n)^{2}/2\tau}\nonumber\\
\rho_{\tau}^{d+2}(\mathbf{a},\mathbf{x})  & =-e^{d\tau/2}\frac{1}{2\pi
\sin\theta}\frac{d}{d\theta}\rho_{\tau}^{d}(\mathbf{a},\mathbf{x}%
).\label{inductive}%
\end{align}
Here $\theta$ is a complex-valued quantity satisfying $\cos\theta
=\mathbf{a}\cdot\mathbf{x}/r^{2}.$ (There is precisely one such $\theta$ with
$0\leq\operatorname{Re}\theta\leq\pi.$) See \cite{HM} for formulas in the
even-dimensional case.

\section{The large $r$ limit}

The purpose of this paper is to study the behavior of the coherent states in
the limit $r\rightarrow\infty.$ For simplicity we consider only the
odd-dimensional case, although the same results almost certainly hold in the
even-dimensional case as well. A region of fixed size $R$ in a sphere of
radius $r$ will look Euclidean as long as $r\gg R.$ So we expect the coherent
states for large $r$ on $S^{d}$ to look like the usual coherent states on
$\mathbb{R}^{d}$ provided that the coherent states are concentrated into a
region of size $R\ll r.$

Now, the spatial size of the coherent states is controlled by the
dimensionless parameter $\tau=\hbar/m\omega r^{2}.$ Specifically, if $\Delta
X$ denotes the approximate spatial width of the coherent states then we expect
that $\Delta X\approx\sqrt{\hbar/2m\omega}$ (as it is in the Euclidean case)
at least if this quantity is small compared to $r.$ In that case we will have
\[
\frac{\Delta X}{r}\approx\frac{\sqrt{\hbar/2m\omega}}{r}=\sqrt{\frac{\tau}{2}%
}.
\]
We will prove that if $\omega,$ $m,$ and $\hbar$ remain fixed and $r$ tends to
infinity then indeed the coherent states have spatial width approximately
equal to $\sqrt{\hbar/2m\omega}$ and that the coherent states become (in a
sense to be described shortly) precisely the usual Gaussian wave packets on
$\mathbb{R}^{d}.$ (See (\ref{lim.form}) and (\ref{lim.form2}) below.) The
proof amounts to analyzing the behavior of the heat kernel for small $\tau,$
since $\tau=\hbar/m\omega r^{2}$ tends to zero as $r$ tends to infinity with
$\omega,$ $m,$ and $\hbar$ fixed.

Kowalski and Rembieli\'{n}ski do not have a parameter comparable to our
$\omega$ in \cite{KR1}. This means that what they do corresponds to the
$\tau=1$ case of our construction. Since the value of $\tau$ is fixed, the
angular dependence of the coherent states in \cite{KR1} is independent of $r,
$ that is, the coherent states in \cite{KR1} simply scale proportionally to
$r.$ (See Equation (5.3) in \cite{KR1}.) It seems to us, therefore, that one
cannot get the canonical coherent states on $\mathbb{R}^{d}$ in the large-$r$
limit without introducing the parameter $\omega.$

Another way of thinking about the significance of $\omega$ is that $m\omega$
has units of momentum divided by position. On the classical side this gives a
way of writing dimensionally correct combinations of position and momentum
(without using $\hbar$), as in (\ref{a.form}). On the quantum side we expect
that the ratio of the width in momentum space to the width in position space
of the coherent states is approximately $m\omega.$

The coherent states are labeled by points $\mathbf{a}$ in the complex sphere,
which are in one-to-one correspondence with points in the classical phase
space by means of (\ref{a.form}). We consider a coherent state where the
position part of the label is at a fixed distance (independent of $r$) from
the north pole, and we evaluate that coherent state at a point that is also at
a fixed distance from the north pole. Specifically, we choose a point
$(\mathbf{x}_{0},\mathbf{p}_{0})$ in $\mathbb{R}^{d}\times\mathbb{R}^{d}$ and
we let $\mathbf{\tilde{x}}_{0}$ be the point in $S^{d}$ given by
\begin{equation}
\mathbf{\tilde{x}}_{0}=\left(  \mathbf{x}_{0},\sqrt{r^{2}-x_{0}^{2}}\right)
,\label{xx}%
\end{equation}
which makes sense for all $r$ with $r^{2}\geq x_{0}^{2}.$ We also let
$\mathbf{\tilde{p}}_{0}$ be given by
\begin{equation}
\mathbf{\tilde{p}}_{0}=\left(  \mathbf{p}_{0},-\frac{\mathbf{p}_{0}%
\cdot\mathbf{x}_{0}}{\sqrt{r^{2}-x_{0}^{2}}}\right)  \label{pp}%
\end{equation}
so that $\mathbf{\tilde{x}}_{0}\cdot\mathbf{\tilde{p}}_{0}=0,$ that is,
$\left(  \mathbf{\tilde{x}}_{0},\mathbf{\tilde{p}}_{0}\right)  \in T^{\ast
}(S^{d})$. We now consider the coherent state $\psi_{\mathbf{a(\tilde{x}}%
_{0}\mathbf{,\tilde{p}}_{0}\mathbf{)}}.$ We then fix another point
$\mathbf{x}$ in $\mathbb{R}^{d},$ and we let $\mathbf{\tilde{x}}%
=(\mathbf{x},(r^{2}-x^{2})^{1/2}).$ (See Figure 1.)
 We will show that $\psi_{\mathbf{a(\tilde
{x}}_{0}\mathbf{,\tilde{p}}_{0}\mathbf{)}}(\mathbf{\tilde{x}}),$ viewed as a
function of $\mathbf{x},$ behaves for large $r$ like a Gaussian wave packet
centered at $\mathbf{x}=\mathbf{x}_{0}$ and with momentum $\mathbf{p}_{0}.$

$$\hbox{\epsfxsize=3.00 in \epsfbox{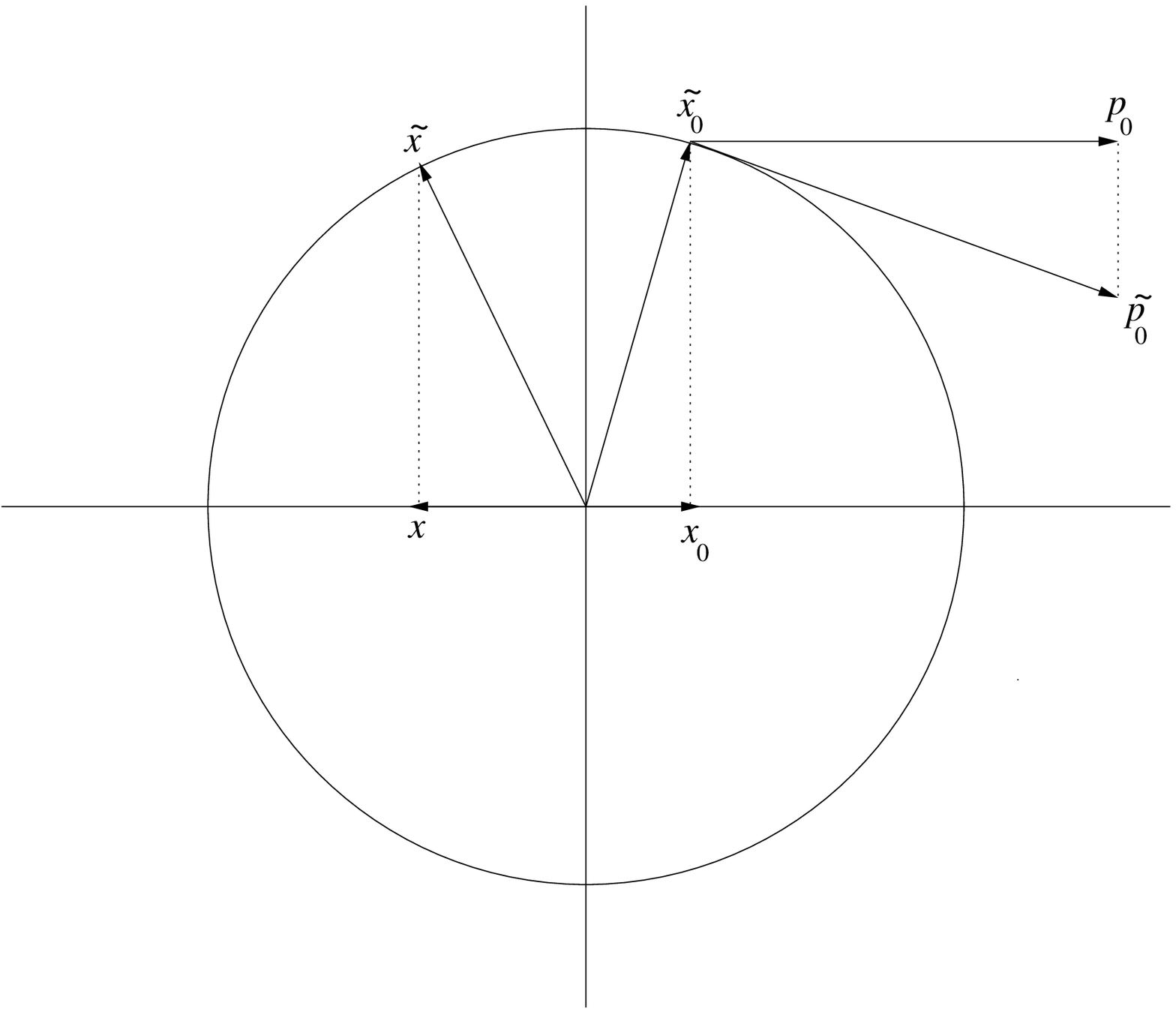}}$$

\centerline{Figure 1}\medskip

Note that for large $r$ we have $\mathbf{\tilde{x}}_{0}\approx(\mathbf{x}%
_{0},r)$ and $\mathbf{\tilde{p}}_{0}\approx(\mathbf{p}_{0},0).$ It follows
from (\ref{a.form}) that for large $r$ we have
\[
\mathbf{a(\tilde{x}}_{0}\mathbf{,\tilde{p}}_{0}\mathbf{)}\approx\left(
\mathbf{x}_{0}+i\frac{\mathbf{p}_{0}}{m\omega},r\right)  .
\]
That is, the first $d$ components of $\mathbf{a(\tilde{x}}_{0}\mathbf{,\tilde
{p}}_{0}\mathbf{)}$ give just $\mathbf{x}_{0}+i\,\mathbf{p}_{0}/m\omega$ as in
the Euclidean case, and the last component of $\mathbf{a(\tilde{x}}%
_{0}\mathbf{,\tilde{p}}_{0}\mathbf{)}$ is just a constant. Heuristically this
means that on the quantum side (when applied to functions supported in a set
of fixed radius $R$ around the north pole) we should have the first $d$
components of $\mathbf{A}$ approximately equal to $X+iP/m\omega$ and the last
component of $\mathbf{A}$ approximately equal to $rI.$ This explains
heuristically why the coherent states (eigenvectors of $\mathbf{A}$) should
look like their Euclidean counterparts in the large $r$ limit.

\begin{theorem}
\label{main}The coherent states on odd-dimensional spheres have the following
limiting property for large $r.$ Fix $\mathbf{x}_{0}$ and $\mathbf{p}_{0}$ in
$\mathbb{R}^{d},$ and let $\mathbf{z}=\mathbf{x}_{0}+i\,\mathbf{p}_{0}%
/m\omega.$ Then
\begin{equation}
\lim_{r\rightarrow\infty}r^{-d}\psi_{\mathbf{a(\tilde{x}}_{0}\mathbf{,\tilde
{p}}_{0}\mathbf{)}}(\mathbf{\tilde{x}})=\left(  \frac{m\omega}{2\pi\hbar
}\right)  ^{d/2}\exp\left\{  -\frac{(\mathbf{z}-\mathbf{x})^{2}}%
{2\hbar/m\omega}\right\}  \label{lim.form}%
\end{equation}
where the limit is uniform for $\mathbf{x}$ in compact subsets of
$\mathbb{R}^{d}.$ Here $\mathbf{\tilde{x}}_{0},$ $\mathbf{\tilde{p}}_{0}$, and
$\mathbf{\tilde{x}}$ are defined by (\ref{xx}) and (\ref{pp}) and
$(\mathbf{z}-\mathbf{x})^{2}$ means $(z_{1}-x_{1})^{2}+\cdots+(z_{d}%
-x_{d})^{2}.$
\end{theorem}

The right side of (\ref{lim.form}) is the usual Gaussian coherent state, which
can also be written as
\begin{equation}
c_{z}\left(  \frac{m\omega}{2\pi\hbar}\right)  ^{d/2}\exp\left\{
-\frac{(\mathbf{x}-\mathbf{x}_{0})^{2}}{2\hbar/m\omega}\right\}
e^{i\mathbf{p}_{0}\cdot\mathbf{x/}\hbar},\label{lim.form2}%
\end{equation}
where $c_{\mathbf{z}}=\exp(-i\mathbf{p}_{0}\cdot\mathbf{x}_{0}/\hbar
)\exp(p_{0}^{2}/2\hbar m\omega).$

\section{Proofs}

Theorem \ref{main} would follow from standard heat kernel asymptotics if the
angle $\theta$ were real. As it is, we need to verify that the expected
behavior of the heat kernel holds for small $\tau$ and small \textit{complex}
$\theta.$ Recall that $\psi_{\mathbf{a(\tilde{x}}_{0}\mathbf{,\tilde{p}}%
_{0}\mathbf{)}}(\mathbf{\tilde{x}})=\rho_{\tau}^{d}\left(  \mathbf{a(\tilde
{x}}_{0}\mathbf{,\tilde{p}}_{0}\mathbf{),\tilde{x}}\right)  $ depends only on
the complex angle between $\mathbf{a(\tilde{x}}_{0}\mathbf{,\tilde{p}}%
_{0}\mathbf{)}$ and $\mathbf{\tilde{x}}$. We will denote this function by
$\rho_{\tau}^{d}(\theta).$ Since the heat kernel $\rho_{\tau}^{d}(\theta)$ is
an even entire function of $\theta,$ and so depends only on $\theta^{2},$ we
need to understand the dependence of $\theta^{2}$ on $\mathbf{z}%
=\mathbf{x}_{0}+i\mathbf{p}_{0}/m\omega$ and $\mathbf{x}.$ The following lemma
provides the answer.

\begin{lemma}
\label{theta}Fix $\mathbf{x}_{0}$ and $\mathbf{p}_{0}$ in $\mathbb{R}^{d}.$
Let $K$ be a compact set in $\mathbb{R}^{d}$ and let $s$ be a positive number
with $s<\pi$. Then for all sufficiently large $r$ and all $\mathbf{x}$ in $K$
there are solutions to $\cos\theta=r^{-2}(\mathbf{a(\tilde{x}_{0},\tilde{p}%
}_{0}\mathbf{)\cdot\tilde{x}})$ with $\left|  \theta\right|  <s$. These
solutions are unique up to a sign and satisfy
\[
\theta^{2}=\frac{(\mathbf{z}-\mathbf{x})^{2}}{r^{2}}+q(r,\mathbf{z}%
,\mathbf{x}),
\]
where $\mathbf{z}=\mathbf{x}_{0}+i\frac{\mathbf{p}_{0}}{m\omega}$ and
\[
|q(r,\mathbf{z},\mathbf{x})|\leq Cr^{-4}%
\]
for a fixed positive constant $C$ and all $\mathbf{x}\in K.$
\end{lemma}

The proof uses only that $\cos\theta\approx1-\theta^{2}/2$ for small $\theta,$
and is omitted. Note that as $r$ tends to infinity both $\tau$ and $\theta$
tend to zero but that $\theta^{2}/2\tau$ tends to $(\mathbf{z}-\mathbf{x}%
)^{2}/(2\hbar/m\omega)$ as a function of $\mathbf{x}$ uniformly on compact
subsets of $\mathbb{R}^{d}.$ This observation is essential to the proof of
Theorem \ref{main}.

\begin{lemma}
\label{remd}For odd positive integers $d,$ let $P_{d}(\tau,\theta)$ be defined
inductively by
\begin{align*}
P_{1}(\tau,\theta)  & =(2\pi\tau)^{-1/2}e^{-\theta^{2}/2\tau}\\
P_{d+2}(\tau,\theta)  & =-\frac{e^{d\tau/2}}{2\pi}\frac{1}{\sin\theta}%
\frac{\partial P_{d}(\tau,\theta)}{\partial\theta}%
\end{align*}
for complex $\theta$ with $\left|  \theta\right|  <\pi.$ Then $P_{d}%
(\tau,\theta)$ is an even holomorphic function of $\theta$ in the set $\left|
\theta\right|  <\pi$ (with a removable singularity at the origin) for all
$d,\tau.$

Let $R_{d}(\tau,\theta)$ be defined so that
\[
\rho_{\tau}^{d}(\theta)=P_{d}(\tau,\theta)+R_{d}(\tau,\theta).
\]
Then there exist positive constants $s_{d},$ $B_{d},$ and $C_{d}$ such that
\[
\left|  R_{d}(\tau,\theta)\right|  \leq B_{d}e^{-C_{d}/\tau}%
\]
for all $\theta$ with $\left|  \theta\right|  <s_{d}$ and all $\tau$ with
$0<\tau<1.$
\end{lemma}

\begin{proof}
The statement concerning $P_{d}(\tau,\theta)$ follows easily by induction.
Therefore, we need only prove the bound on $R_{d}(\tau,\theta).$ In light of
the formula for $\rho_{\tau}^{1}$ we have
\[
R_{1}(\tau,\theta)=\frac{1}{\sqrt{2\pi\tau}}\sum_{n\neq0}e^{-\frac
{(\theta-2\pi n)^{2}}{2\tau}}.
\]
If we take, say, $s_{1}=\frac{\pi}{2},$ then $\operatorname{Re}(\theta-2\pi
n)>3\pi/2$ for all non-zero integers $n$ and all $\theta$ with $|\theta
|<s_{1}.$ From this it easily follows that, say,
\[
|R_{1}(\tau,\theta)|\leq Be^{-\pi^{2}/2\tau}%
\]
for some $B$ and for all $\theta$ with $|\theta|<s_{1}.$ This establishes the
desired result in the $d=1$ case.

Assume now we have the desired bound for $R_{d}(\tau,\theta).$ In light of the
inductive formula (\ref{inductive}) for the heat kernel and the definition of
$P_{d}$ we have
\[
R_{d+2}(\tau,\theta)=-\frac{e^{d\tau/2}}{2\pi}\frac{1}{\sin\theta}%
\frac{\partial R_{d}}{\partial\theta}(\tau,\theta)
\]
which we write as
\begin{equation}
R_{d+2}(\tau,\theta)=-\frac{e^{d\tau/2}}{2\pi}\frac{\theta}{\sin\theta}%
\frac{1}{\theta}\frac{\partial R_{d}}{\partial\theta}(\tau,\theta).\label{rd}%
\end{equation}
Set $s_{d+2}$ equal to any positive number strictly less than $s_{d},$ and let
$R_{d}(\tau,\theta)=\sum_{n=0}^{\infty}a_{n}(\tau)\theta^{2n}$ be the power
series expansion for $R_{d}(\tau,\theta)$. By the Cauchy estimates, we have
\[
|a_{n}(\tau)|\leq B_{d}e^{-C_{d}/\tau}s_{d}^{-2n}%
\]
for all $0<\tau<1.$ Then for $\left|  \theta\right|  <s_{d+2}$ we have
\[
\frac{1}{\theta}\frac{\partial R_{d}}{\partial\theta}(\tau,\theta)=\sum
_{n=1}^{\infty}2n\,a_{n}(\tau)\theta^{2n-2}.
\]
Therefore, for $\left|  \theta\right|  <s_{d+2}$ and $0<\tau<1$ we have
\begin{align*}
\left|  \frac{1}{\theta}\frac{\partial R_{d}}{\partial\theta}(\tau
,\theta)\right|    & \leq B_{d}e^{-C_{d}/\tau}\sum_{n=1}^{\infty}%
2n\frac{|\theta|^{2n-2}}{s_{d}^{2n-2}}\\
& \leq B_{d}e^{-C_{d}/\tau}\sum_{n=1}^{\infty}2n\frac{s_{d+2}^{2n-2}}%
{s_{d}^{2n-2}}=De^{-C_{d}/\tau}%
\end{align*}
for a positive constant $D.$ The bound for $R_{d+2}(\tau,\theta)$ now follows
from (\ref{rd}).
\end{proof}

We use the notation of the previous lemma in what follows.

\begin{lemma}
\label{domd} For all odd positive integers $d$, there exist positive numbers
$s_{d}<\pi$ such that
\[
\lim_{\tau\rightarrow0^{+}}\frac{P_{d}(\tau,\theta)}{(2\pi\tau)^{-d/2}%
e^{-\theta^{2}/2\tau}}=\left(  \frac{\theta}{\sin\theta}\right)  ^{(d-1)/2}%
\]
uniformly on compact subsets of the set $\left|  \theta\right|  <s_{d}.$
\end{lemma}

It is worth noting that the function on
the right in the lemma is $J^{-1/2}(\theta)$ where $J$ is the Jacobian of the
exponential mapping for $S^{d},$ even though we will not make use of this fact.

\begin{proof}
This is trivially true for $d=1.$ We proceed by induction. Define $g_{d}%
(\tau,\theta)$ so that
\[
P_{d}(\tau,\theta)=(2\pi\tau)^{-d/2}e^{-\theta^{2}/2\tau}g_{d}(\tau,\theta)
\]
and observe that $g_{d}(\tau,\theta)$ is even and holomorphic on
$|\theta|<s_{d}.$ Then
\begin{align}
P_{d+2}(\tau,\theta)  & =-\frac{e^{d\tau/2}}{2\pi}\frac{1}{\sin\theta}%
\frac{\partial P_{d}}{\partial\theta}(\tau,\theta)\nonumber\\
& =(2\pi\tau)^{-(d+2)/2}e^{-\theta^{2}/2\tau}\left\{  e^{d\tau/2}\frac{\theta
}{\sin\theta}g_{d}(\tau,\theta)-\tau e^{d\tau/2}\frac{\theta}{\sin\theta}%
\frac{1}{\theta}\frac{\partial g_{d}}{\partial\theta}(\tau,\theta)\right\}
.\label{p.est}%
\end{align}
Our induction hypothesis on $P_{d}(\tau,\theta)$ implies that $g_{d}%
(\tau,\theta)$ tends uniformly to the function $(\theta/\sin\theta)^{(d-1)/2}$
on compact subsets of $|\theta|<s_{d}$ as $\tau\rightarrow0^{+}.$ It follows
by basic complex analysis that $(1/\theta)\partial g_{d}(\tau,\theta
)/\partial\theta$ tends to $1/\theta$ times the derivative of $(\theta
/\sin\theta)^{(d-1)/2}$ uniformly on compact sets. Thus, because of the factor
of $\tau,$ the second term in brackets in (\ref{p.est}) tends to zero
uniformly on compact sets. The first term in the brackets tends to
$(\theta/\sin\theta)^{(d+1)/2}$ uniformly on compact sets as $\tau
\rightarrow0^{+},$ which establishes the lemma for dimension $d+2.$
\end{proof}

We now put our results together to supply the proof of Theorem \ref{main}.

\begin{proof}
Recall that $\tau=\hbar/m\omega r^{2}$ and that we must evaluate the heat
kernel at a value of $\theta$ where $\cos\theta=r^{-2}(\mathbf{a(\tilde{x}%
}_{0}\mathbf{,\tilde{p}}_{0}\mathbf{)\cdot\tilde{x}}).$ Lemma \ref{theta}
shows that we may choose $\theta$ satisfying $|\theta|<s_{d}$ if $r$ is
sufficiently large$.$ Therefore, by Lemma \ref{remd}, we need only calculate
the limit of $r^{-d}P_{d}(\hbar/m\omega r^{2},\theta)$ as $r\rightarrow
\infty.$ As noted above,
\begin{equation}
\lim_{r\rightarrow\infty}\frac{\theta^{2}}{2\tau}=\frac{(\mathbf{z}%
-\mathbf{x})^{2}}{2\hbar/m\omega}\label{thetalim}%
\end{equation}
uniformly for $\mathbf{x}\in K.$ Therefore, using (\ref{thetalim}), Lemma
\ref{domd} with the fact that we may choose $\theta^{2}$ to tend uniformly to
zero as $r\rightarrow\infty,$ and the fact that continuous functions are
uniformly continuous on compact subsets, it follows that
\[
\lim_{r\rightarrow\infty}r^{-d}P_{d}\left(  \frac{\hbar}{m\omega r^{2}}%
,\theta\right)  =\left(  \frac{m\omega}{2\pi\hbar}\right)  ^{d/2}\exp\left\{
-\frac{(\mathbf{z}-\mathbf{x})^{2}}{2\hbar/m\omega}\right\}
\]
uniformly on $K.$
\end{proof}


\begin{thebibliography}{9}
\bibitem{H1}B. Hall, The Segal--Bargmann ``coherent state'' transform for
compact Lie groups, \textit{J. Funct. Anal}. \textbf{112} (1994), 103-151.

\bibitem {H2}B. Hall, The inverse Segal--Bargmann transform for compact Lie
groups, \textit{J. Funct. Anal.} \textbf{143} (1997), 98--116.

\bibitem {H3}B. Hall, Harmonic analysis with respect to heat kernel measure,
\textit{Bull. Amer. Math. Soc. (N.S.)} \textbf{38} (2001), 43--78.

\bibitem {HM}B. Hall and J. Mitchell, Coherent states on spheres, \textit{J.
Math. Phys. }\textbf{43} (2002), 1211-1236.

\bibitem {KR1}K. Kowalski and J. Riembeli\'{n}ski, Quantum mechanics on a
sphere and coherent states, \textit{J. Phys. A} \textbf{33} (2000), 6035-6048.

\bibitem {St}M. Stenzel, The Segal--Bargmann transform on a symmetric space of
compact type, \textit{J. Funct. Anal}. \textbf{165} (1999), 44--58.

\bibitem {T}T. Thiemann, Reality conditions inducing transforms for quantum
gauge field theory and quantum gravity, \textit{Classical Quantum Gravity}
\textbf{13} (1996), 1383--1403.
\end{thebibliography}
\end{document}